\documentclass[submission]{raa}            

\usepackage{graphicx,times}             %for PS/EPS graphics inclusion, new
\usepackage[english]{babel}
\usepackage{natbib}
\usepackage{graphicx,epsfig}
\usepackage{color,subfigure}
\usepackage{mwe}
\usepackage{morefloats}
\usepackage{amsmath,amssymb}
\usepackage{lscape}
\usepackage{hyperref}
\usepackage{url}
\usepackage{fancyhdr}

\begin{document}

   \title{UV and X-Ray Variability of the Narrow-Line Seyfert~1 Galaxy Ark~564
%\,$^*$
%\footnotetext{$*$ Supported by the National Natural Science Foundation of China.}
}
%   \subtitle{I. Place Your Subtitle Here}

   \volnopage{Vol.0 (200x) No.0, 000--000}      %%preserved for Editor. DOn't remove!
   \setcounter{page}{1}          %%starting page, preserved for Editor. DOn't remove!

   \author{Savithri H. Ezhikode
      \inst{1}
   \and Gulab C. Dewangan
      \inst{2}
   \and Ranjeev Misra
      \inst{2}
   \and Shruti Tripathi
      \inst{3}
   \and Ninan Sajeeth Philip
      \inst{1}
   \and Ajit K. Kembhavi
      \inst{2}   
   }
%% Here is an example of three authors come from different institutes.
%% For single author or all the authors from an institute, use "\inst{}" only

   \institute{Department of Physics, St. Thomas College, Kozhencherry, Kerala, India\\
%% Please give the E-mail address of the author, to whom future correspondence and
%% offprint requests will be sent.
        \and
             Inter-University Centre for Astronomy \& Astrophysics, Post Bag 4, Ganeshkhind, Pune, India\\
        \and
             Physics Department, Bishop’s University, 2600 College Street, Sherbrooke, QC, Canada J1M 1Z7\\
   }

   \date{Received~~2009 month day; accepted~~2009~~month day}

\abstract{ We analyse eight \textit{XMM-Newton} observations of the bright Narrow-Line
Seyfert~1 galaxy Arakelian 564 (Ark~564). These observations, separated by
$\sim 6$~days, allow us to look for correlations between the simultaneous
UV emission (from the Optical Monitor) with not only the X-ray flux but also
with the different X-ray spectral parameters. The X-ray spectra from all the
observations are found to be adequately fitted by a double Comptonization model
where the soft excess and the hard X-ray power law are represented by thermal
Comptonization in a low temperature plasma and hot corona, respectively. 
Apart from the fluxes of each component, the hard X-ray power law index 
is found to be variable. These results suggest that the variability is associated with 
changes in the geometry of the inner region. The UV emission 
is found to be variable and well correlated with the high energy index while 
the correlations with the fluxes of each component are found to be weaker.
Using viscous time-scale arguments we rule out the possibility that the UV variation
is due to fluctuating accretion rate in the outer disc. If the UV variation is driven
by X-ray reprocessing, then our results indicate that the strength of the X-ray
reprocessing depends more on the geometry of the X-ray producing inner region rather than
on the X-ray luminosity alone.
}

   \authorrunning{Ezhikode et. al. }            %author_head in even pages
   \titlerunning{UV and X-Ray Variability of the NLS~1 Ark~564 }  % title_head in odd pages

   \maketitle
%% The author head (on even pages) and the title head (on odd pages) will be
%% automatically extracted from \author{} and \title{}. Whenever the title is too long,
%% you will be asked to supply a shorter one by inserting either \authorrunning{} or
%% \titlerunning{} before \maketitle. Anyway, you can specify your own heads.
%%
%%
%% Note: In the following text body of your manuscript, please note several differences from
%%       other major journals:
%% (1) \subsection{Please Capitalize the First Letter of Each Notional Word in Subsection Title}
%% (2) Please Capitalize the First Letter of Each Notional Word in all tables' captions

%
%________________________________________________ sections below
%
\section{Introduction}           %% first-level sections will be auto-capitalized
\label{sect:intro}

Active Galactic Nuclei (AGN) emit over a wide range of the electromagnetic
spectrum and their spectra show strong optical/UV emission lines which are not
present in the spectrum of a normal galaxy. AGN are believed to harbour a
supermassive black hole (SMBH) of mass $\sim10^6-10^9M_{\odot}$. The accretion 
of matter on to the SMBH is the major source of radiation in AGN, and they can outshine
the stellar emission of the host galaxy (e.g. \citealt{Peterson,Volker}). 
According to the standard model of AGN, the matter accreted from the host galaxy 
forms an accretion disc surrounding the central black hole and the spectrum 
emitted from the disc peaks in the optical/UV band. Furthermore, there is a hot corona 
above the disc which inverse Compton scatters the disc photons resulting in X-ray emission. 
The broadband X-ray emission is one of the fundamental characteristics defining AGN (e.g. \cite{Volker}).

A significant property of AGN is their continuum variability over the entire
electromagnetic spectrum. AGN, in general, show strong X-ray variability and
a subset of AGN called Narrow Line Seyfert~1 (NLS1) shows extreme variability (e.g. \cite{Boller96}).
Relationship between emission in different bands provides important insights
into the nature of AGN. A number of previous studies have shown that X-ray
and optical/UV variations in AGN are well correlated \citep{Nandra_etal1998, Edelson96, Smith_Vaughan07,McHardy14}. 
There are two basic models to explain the UV variability in AGN: 
the UV variability could be due to accretion rate variation in the outer disc 
or it could be due to X-ray reprocessing (e.g. \cite{McHardy14}).
In the first case, the UV flux variation can provide information about accretion rate
variation while in the second case, the correlation between the soft/hard
X-ray component and the UV emission can reveal which X-ray spectral component
has the greater affect on the outer disc.

A multiwavelength campaign undertaken by \citet{Edelson96} observed a strong correlation
between the X-ray, UV and optical variability in the Seyfert~1 (Sy~1) galaxy NGC 4151, with
no detectable lags. The UV observations were taken with a sampling interval of $\sim$0.05~d
while the X-ray observations were taken twice per day. The obtained results suggest that UV
emission in the source is produced by the reprocessing of primary X-rays. A recent study by
\cite{McHardy14} investigates the relationship between the X-ray, UV and optical variability
of the Sy~1 galaxy NGC 5548. They analysed 554 \textit{Swift} (XRT $\&$ UVOT) observations of
the source, typically taken every two days, over a period of 750~d. The study strengthens the
short time-scale correlations between X-ray and UV/optical bands and the lag measurements of
this object also lead to the conclusion that UV/optical variability is due to the reprocessing
of X-rays. \cite{Shemmer03} studied the X-ray$-$optical correlation of the NLS1 galaxy NGC 4051
based on the data available from 2000 May-July observations. The optical data were retrieved from
\textit{Wise Observatory} and X-ray data from \textit{RXTE}. They obtained \textit{RXTE} data from
251 observations with an interval of 6~h. They conclude that the observed X-ray$-$optical correlation
in the source can be explained as a combined effect of X-ray reprocessing and the propagation of
perturbations from the outer disc to the X-ray emitting region.

In this work, we study the source Arakelian 564 (Ark~564) which is an X-ray bright NLS~1 \citep{Brandt94,Vaughan99} found in the nearby universe with redshift z = 0.0247. It is a
well studied source which is known to accrete at super-Eddington rate \citep{Mullaney09}.
Earlier studies have shown that the high energy spectrum of Ark~564 is characterised by a
steep power law \citep{Vignali04,Matsumoto04}. \citet{Vignali04} detected absorption
corresponding to the O VII K-edge ($\sim$ 0.73~keV) in two different \textit{XMM-Newton}
observations (2000 \& 2001) of the source. They also obtained evidence for significant
X-ray variability of the object both at low and high energies. The spectral variability
analysis of Ark~564 was carried out by \citet{Brinkman07} with the longest exposure observation
available from \textit{XMM-Newton}. They found that both the soft and hard X-ray flux are
highly variable on short time-scales. The cross-correlation analysis of light curves showed
some delay in the observation of hard band photons with respect to soft band photons. The
high energy photon index $\Gamma$ was also variable and it was found to be leading the variations
in soft and hard energy bands. A previous study using ASCA observations by \citet{Bian_zhao03}
determined the relations between hard X-ray variability, photon index $\Gamma$ and Eddington ratio
$\dot{m}$ of a sample of AGN, including Ark~564. They found that the X-ray variability and
central black hole mass $M_{BH}$ are strongly anti-correlated while a weak correlation exists
for the variability and $\dot{m}$. This means that small value of $M_{BH}$ is responsible for the
variability of NLS1s. The study also discovered a strong correlation between $\Gamma$ and $\dot{m}$.

\citet{Smith_Vaughan07} examined the X-ray and optical variability of Ark~564 and seven
other Sy~1 galaxies over a period of $\sim$1~d. The source was variable in X-rays, but
not in the optical band. But another study by \cite{Shemmer01} investigated the optical$-$UV$-$X-ray
connection of Ark~564 over a longer period of time of $\sim$50~d. It was a two year long
multiwavelength monitoring program in which the X-ray observations were covered with \textit{RXTE}
and \textit{ASCA} while the UV observations with \textit{HST}. They observed a significant
correlation of the continua, where the X-ray continuum was followed by UV with a time lag of 0.4~d
which in turn was followed by the optical band by $\sim$2~d. The soft X-ray flux was also found to be
well correlated with the hard band flux with zero time lag. Rapid X-ray variations of $\lesssim$1~d
was observed in the source, but the mean flux was constant on time-scales $>$30~d.

A study has been carried out by \cite{GCD_Ark564_07} of the soft excess emission from the NLS1
galaxies Ark~564  and Mrk 1044. They argued that the soft excess emission from Ark~564 can be
explained by considering a two-component corona in the source. According to their model, the
geometry of the corona is such that it consists two different physical regions, one being
optically thick and cool while the other a high-temperature, optically thin region. The hot
corona extends above the low-temperature corona while the latter is coupled to the inner part
of the accretion disc. The optical/UV photons emitted from the accretion disc are Comptonized
by the optically thick corona leading to the soft X-ray emission. The geometry suggests that
a fraction of these scattered photons again get inverse Compton scattered by the optically thin
corona giving rise to the hard X-ray spectrum.

In this work, we examine the correlation between the X-ray and UV emission from Ark~564, by
using eight \textit{XMM-Newton} observations in 2011. The data from these observations have already been
analysed to study the observed time lag between the soft and hard X-ray emissions \citep{Legg12}
and the frequency-dependent Fe K lags \citep{Kara13}. \cite{Legg12} detected a delayed
($\sim 1000s$) hard X-ray emission in the 4$-$7.5~keV with respect to a flaring in the soft
X-ray band (0.4$-$1~keV). In view of these results \cite{Giustini2015} reported on the X-ray
spectral properties using \textit{XMM-Newton} and \textit{Suzaku} observations of Ark~564 by
analysing the time-averaged, flux-selected and time-resolved spectra. They interpreted the
delayed hard excess component as the reprocessing of soft photon flares Compton up-scattered
in a medium situated at 10$-$100 of gravitational radii. However these studies have been limited
to the variations in the X-ray band only. Moreover, most of the previous UV/X-ray correlation studies
of the source have concentrated on the variation of total X-ray counts or flux with that of the
UV band. The high quality X-ray spectral data available by \textit{XMM-Newton} provide us the
opportunity to study the variation of different X-ray spectral components with the UV flux. In the
present work, we fit the spectra utilising some physical models to obtain the spectral parameters
and then study the correlation between these parameters and the UV flux. 

The paper is organized as follows. In Section 2, we describe the observations and data reduction.
We discuss the variability of the source and spectral analysis in Section 3 and Section 4, respectively.
Then we present the correlations in Section 5. Finally, we summarize and discuss our results in Section 6.

\section{Observations and Data Reduction}

We used eight observations of Ark~564 taken by \textit{XMM-Newton} between
May and July 2011. The details of these pointed observations are given in
Table~\ref{tab_obs_dtls}. \textit{XMM-Newton} has
simultaneous X-ray and UV exposures for all these observations. X-ray
data from the European Photon Imaging Camera (EPIC) pn and MOS and
UV data from the Optical Monitor (OM) were retrieved from the HEASARC
archive. The EPIC-pn camera \citep{Struder01} is positioned such that
the incoming radiation from the source enters the primary focus
unobstructed, but the EPIC-MOS cameras \citep{Turner01} can receive
only half of the radiation. The EPIC-pn has large effective area at
high energies as well as high quantum efficiency. The pn CCDs are also
less susceptible to pile-up during the observations of bright
sources. We therefore focus our analysis on the pn data, where the
camera was operated in small window mode using a thin filter. OM
observations were performed with the \textit{UVW2} filter in the imaging
mode. The data reduction is done with SAS version 14.0 using the updated 
calibration files available in July 2015.

The event lists for EPIC-pn are filtered for single and double
(PATTERN=4) best quality (FLAG=0) events in the energy range 0.2$-$10~keV. 
To examine the flaring particle background, 
light-curves are extracted in the 10$-$12~keV band for single events. 
The intervals of flaring background are then removed from the event list 
using the threshold rate of 0.1 counts/s obtained from the light curve. 
The source spectra were extracted from circular regions of radius 
36\arcsec around the centre of the source. The background spectra were also extracted 
from two circular regions of 30\arcsec radii on the same chip, but devoid of source photons.
We checked whether any of the observations were affected
by pile-up using the SAS task \textit{epatplot}. But no significant deviation 
in the pattern distribution was observed in any of them. Then we 
rebinned the data with the tool \textit{specgroup}. While binning we ensured that 
each bin has a minimum number of 20 counts. Also the oversampling factor was set to 5 such that 
there are no more than 5 bins to cover the energy resolution.

All the Optical Monitor observations were made with \textit{UVW2}
filter in the Imaging mode. The SAS task `omichain' is used to
reprocess the OM  data, which automatically produces the combined
source list of all the filters. But there is a chance that the
detection algorithm (`omdetect') may misidentify the source as an
extended one, yielding incorrect values of count rate. In order to
avoid this, omichain is run with the option `omdetectdetectextended=no'
for the detection and photometry to be performed as on a point
source. Then the count rate of the source is obtained 
form the combined source list of each observation.

\begin{table*}
\begin{center}
\caption{List of \textit{XMM-Newton} Observations of Ark~564.}
\begin{tabular}{ccccccc}
\hline
\hline
\\
\multicolumn{1}{c}{Observation} & \multicolumn{1}{c}{Observation ID} & \multicolumn{1}{c}{Start Date} & \multicolumn{1}{c}{Duration} & \multicolumn{1}{c}{PN Exposure Time$^{1}$} & \multicolumn{1}{c}{Number of OM}
\\
\multicolumn{1}{c}{Number} & \multicolumn{1}{c}{} & \multicolumn{1}{c}{} & \multicolumn{1}{c}{s} & \multicolumn{1}{c}{s} & \multicolumn{1}{c}{Exposures}\\ \hline
\\
1 & 0670130201 & 2011 May 24  & 59520 & 41180 & 39 \\
2 & 0670130301 & 2011 May 30  & 55919 & 36510 & 40 \\
3 & 0670130401 & 2011 June 05 & 63582 & 31250 & 45 \\
4 & 0670130501 & 2011 June 11 & 67312 & 43850 & 45 \\
5 & 0670130601 & 2011 June 17 & 60919 & 35570 & 45 \\
6 & 0670130701 & 2011 June 25 & 64439 & 29470 & 43 \\
7 & 0670130801 & 2011 June 29 & 58216 & 40500 & 40 \\
8 & 0670130901 & 2011 July 01 & 55915 & 38400 & 40 \\
\hline
\end{tabular}
\label{tab_obs_dtls}
\begin{center}
${^1}${Net exposure time for the EPIC-pn camera}
\end{center}
\end{center}
\end{table*}

\begin{table*}
\tabcolsep 3.0pt
\small
\begin{center}
\caption{X-ray count rate from EPIC-pn in the 0.3$-$10~keV band (Column 2), UV count rate from OM \textit{UVW2} filter (Column 3) and the corresponding UV flux (Column 4) }
\begin{tabular}{cccc}
\hline
\hline
\\
\multicolumn{1}{c}{Observation} & \multicolumn{1}{c}{EPIC-pn Count Rate} & \multicolumn{1}{c}{UVW2 Count Rate} & \multicolumn{1}{c}{F$_{UV}$}
\\
\multicolumn{1}{c}{Number} & \multicolumn{1}{c}{counts s$^{-1}$} & \multicolumn{1}{c}{counts s$^{-1}$} & \multicolumn{1}{c}{10$^{-15}$ erg cm$^{-2}$ s$^{-1}$ \AA$^{-1}$}   \\ \hline
\\

1 & 55.37  $\pm 0.04$ & 1.29 $\pm 0.01$ & 7.39 $\pm 0.04$   \\
2 & 36.33  $\pm 0.03$ & 1.31 $\pm 0.01$ & 7.49 $\pm 0.04$   \\
3 & 37.11  $\pm 0.03$ & 1.22 $\pm 0.01$ & 6.99 $\pm 0.04$  \\
4 & 43.52  $\pm 0.03$ & 1.29 $\pm 0.01$ & 7.37 $\pm 0.04$ \\
5 & 39.78  $\pm 0.03$ & 1.25 $\pm 0.01$ & 7.13 $\pm 0.04$   \\
6 & 24.20  $\pm 0.03$ & 1.18 $\pm 0.01$ & 6.71 $\pm 0.04$ \\
7 & 38.28  $\pm 0.03$ & 1.28 $\pm 0.01$ & 7.30 $\pm 0.04$   \\
8 & 50.91  $\pm 0.04$ & 1.33 $\pm 0.01$ & 7.57 $\pm 0.04$\\
\hline
\end{tabular}
\label{tab_UV_X-ray}
\end{center}
\end{table*}

\begin{figure}
\begin{center}
\includegraphics[trim=0cm 0cm 0cm 0.0cm, clip=true, width=9.0cm, angle=0]{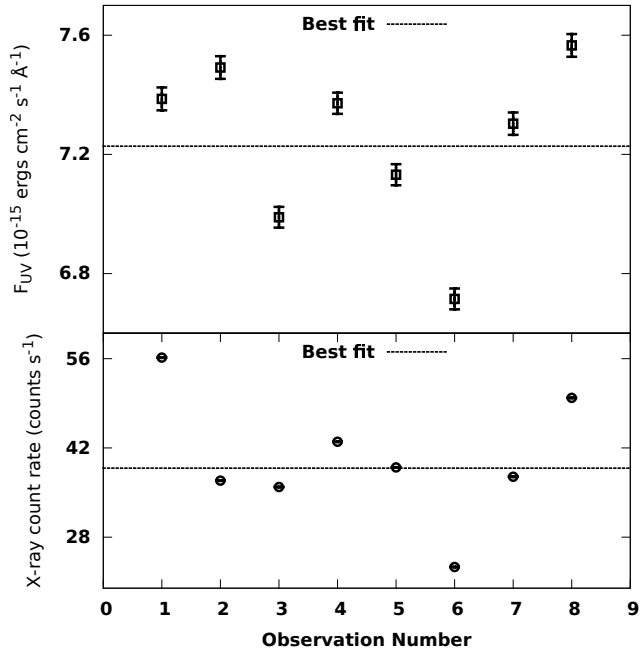}
\caption{The variation in UV flux F$_{UV}$ obtained from \textit{UVW2} filter of Optical Monitor (Upper panel) and X-ray count rate from EPIC-pn in the range 0.3$-$10~keV (Lower panel). The horizontal dot-dashed line in each panel corresponds to the constant best-fit value obtained using $\chi^2$ analysis.The vertical error-bars correspond to $1\sigma$ errors and in case of the X-ray count rate these are very small.}
\label{Fig_UV_Xray_count}
\end{center}
\end{figure}

\begin{figure}
\begin{center}
\includegraphics[trim=0cm 0cm 0cm 0.0cm, clip=true, width =9.0cm, angle=0]{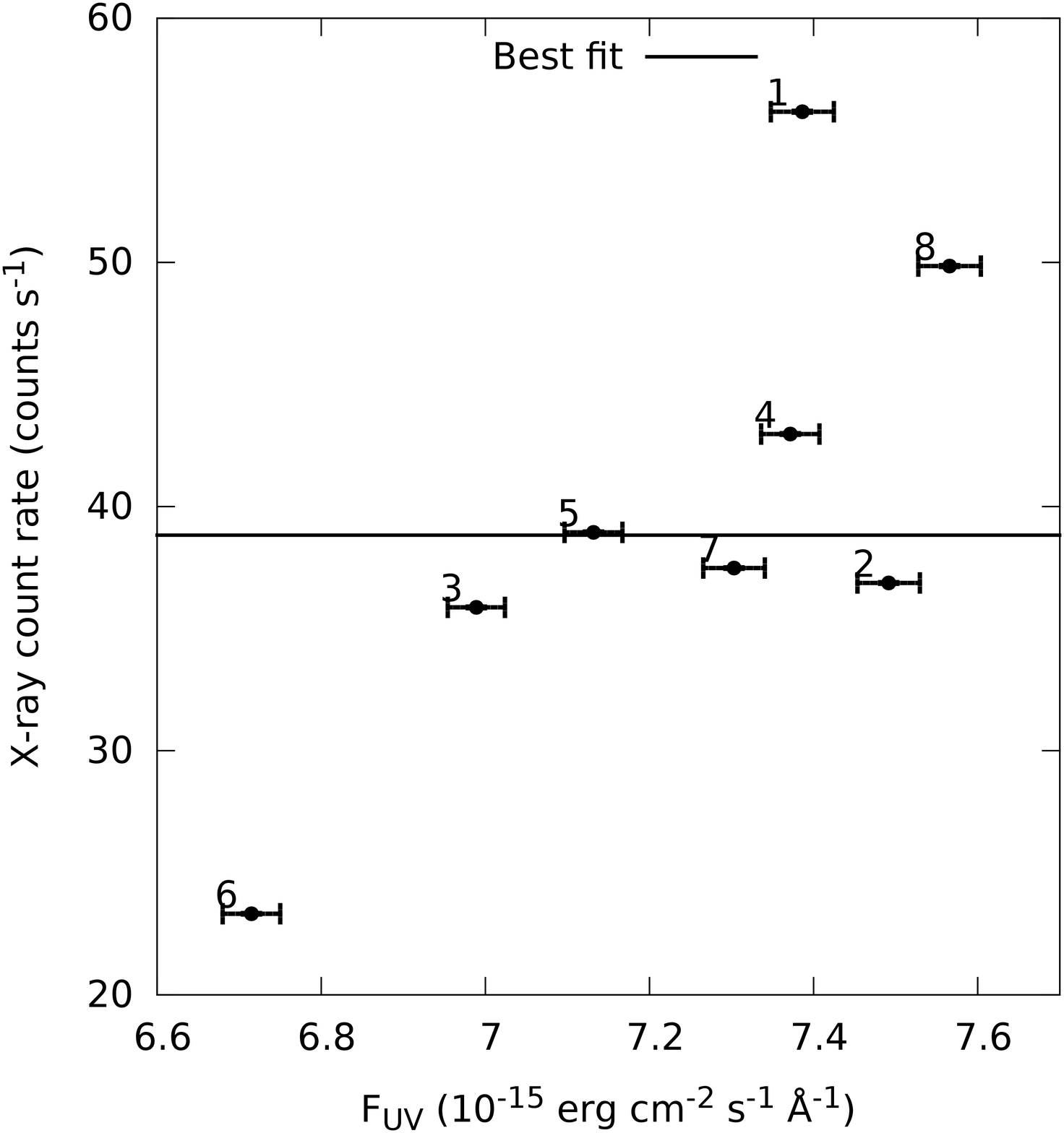}
\caption{The variation in X-ray count rate from EPIC-pn in the range 0.3$-$10~keV. Here, the vertical error-bars ($1\sigma$) are very small. The horizontal solid line corresponds to the constant best-fit value obtained using $\chi^2$ analysis.}
\label{Fig_UV_Xray_var}
\end{center}
\end{figure}

\section{UV $-$ X-ray Variability}

We calculated the UV flux F$_{UV}$ of the source from each 
observation by multiplying the count rate, obtained from the combined source list,
with the conversion factor 5.71$\times$ 10$^{-15}$ erg cm$^{-2}$ s$^{-1}$ \AA$^{-1}$ 
for the \textit{UVW2} filter\footnote{See \textit{XMM-Newton} Optical and UV Monitor (OM) 
Calibration Status document, Talavera 2011 (\url{http://xmm2.esac.esa.int/docs/documents/CAL-TN-0019.pdf})}. Both EPIC-pn and \textit{UVW2} count rates and \textit{UVW2} flux values
are given in Table~\ref{tab_UV_X-ray}. We checked the variability of
the source by $\chi^2$ analysis, which measures the deviation of the
data points from the best-fit constant. 
The analysis yielded large $\chi^2$ values for F$_{UV}$ and pn count rate (0.3$-$10~keV), proving that the source is highly variable in X-ray and UV bands. This is clear from Fig.~\ref{Fig_UV_Xray_count} and Fig.~\ref{Fig_UV_Xray_var}.

In order to confirm that the observed variability is not an artifact
of the instrument we checked the variability in F$_{UV}$ of other
sources which happened to be in the same field of view of all
\textit{UVW2} observations. Among these 23 sources five were found to be
varying in flux while other sources didn't show any sign of variability.
Some of the non-variable sources observed in the same field of view are
NVSS J224244+293856, NVSS J224252+294533, 2MASX J22425351+2943125 etc.

We have also calculated the fractional rms variability amplitude of
both F$_{UV}$ and pn count rate of Ark~564 and the respective values are
$\sim$0.039 and $\sim$0.245. This shows that the observed variance in
X-ray count rate is large compared to the UV flux variance.

\section{Spectral Analysis}

The spectral analysis of EPIC-pn and \textit{UVW2} data was done with 
the XSPEC package (Version 12.8.2). The $\chi^2$ statistic was applied for the spectral 
fitting and the errors calculated  for each parameter correspond to the 90\% confidence range, 
unless otherwise stated.

The template spectral file and response files for the OM data were obtained from the \textit{XMM-Newton} website\footnote{\url{http://xmm2.esac.esa.int/external/xmm_sw_cal/calib/om_files.shtml}}. We used the template file and the measured count rate to create the spectral file for the analysis of \textit{UVW2} data.

As an example of how we have systematically done the spectral analysis we report here 
the analysis of Observation Number 2 (Observation ID: 0670130301). The same analysis 
scheme was applied to the other observations. We started the spectral 
analysis by fitting the EPIC-pn data, in the energy range 3 to 10 keV, using \textit{powerlaw}
together with \textit{TBabs}, the Tuebingen-Boulder ISM absorption model \citep{Wilms2000}. 
This model with photon index $\Gamma \sim 2.3 - 2.5$ and the Galactic column density $N_{H}$ 
fixed to $5.41\times 10^{20} cm^{-2}$ \citep{LAB2005} provides a reasonable fit for all observations. 
It gave a $\chi^2$ of 239.66 for 162 degrees of freedom (dof) for the second observation. 
Some excess emission features were detected in the range 6.4$-$7~keV which might be attributed to the fluorescent Fe emission line. An improvement in fit, $\Delta\chi^2=-24.29$, was observed 
in the second observation, when a red-shifted Gaussian profile (\textit{zgauss}) at $\sim$~6.6~keV 
was included. The line appeared to be broad having a width, $\sigma$=0.25$_{-0.11}^{+0.61}$~keV. 
But this broad Gaussian did not fit properly the spectrum of any of the other observations. 
When we fixed $\sigma$ of the line to 0.5~keV, all observations could achieve a good fit statistic.

\begin{landscape}
\begin{table}
\caption{Best-fit parameters for the model \textit{uvred$\times$zedge$\times$zedge$\times$TBabs}(\textit{ezdiskbb+Simpl$\otimes$Nthcomp+zgauss}) fitted to the \textit{UVW2} and EPIC-pn spectral data}
\begin{center}
\begin{tabular}{lllllllllll}
\hline
\hline
\\
\multicolumn{2}{c}{Model~~~~~~~~~} & \multicolumn{8}{c}{Observation Number} \\ 
\\
\hline
\\
\multicolumn{1}{c}{Component} & \multicolumn{1}{l}{Parameter} & \multicolumn{1}{c}{1} & \multicolumn{1}{c}{2} & \multicolumn{1}{c}{3} & \multicolumn{1}{c}{4} & \multicolumn{1}{c}{5} & \multicolumn{1}{c}{6} & \multicolumn{1}{c}{7} & \multicolumn{1}{c}{8} \\
\hline
\\
\textit{zedge} 1 & E$_{edge}$ (keV) & 0.72 $\pm0.01$
& 0.69 $\pm0.01$
& 0.71 $\pm0.01$
& 0.71 $\pm0.01$
& 0.71 $\pm0.01$
& 0.7 $\pm0.01$
& 0.71 $\pm0.01$
& 0.71 $\pm0.01$
\\\\
& $\tau^{(1)}_{max}$ &  0.11 $\pm0.01$
& 0.12 $^{+0.01 }_{-0.02 }$
& 0.12 $\pm0.02$
& 0.15 $\pm0.02$
& 0.11 $\pm0.02$
& 0.18 $\pm0.03$
& 0.15 $\pm0.02$
& 0.12 $\pm0.02$
\\\\
\textit{zedge} 2 & E$_{edge}$ (keV) & 0.5 $\pm0.01$
& 0.51 $\pm0.02$
& 0.53 $\pm0.02$
& 0.54 $\pm0.01$
& 0.52 $\pm0.01$
& 0.54 $^{+0.01 }_{-0.02 }$
& 0.536 $^{+0.008 }_{-0.003 }$
& 0.52 $\pm0.01$
\\\\
& $\tau^{(1)}_{max}$ & 0.1 $\pm0.02$
& 0.05 $\pm0.02$
& 0.05 $\pm0.03$
& 0.09 $_{- 0.02 }^{+ 0.03 }$
& 0.07 $_{- 0.02 }^{+ 0.03 }$
& 0.11 $\pm0.04$
& 0.12 $\pm0.03$
& 0.06 $\pm0.02$
\\\\
\textit{Simpl} & $\Gamma_{Simpl}$ & 2.60 $\pm0.01$
& 2.56 $\pm0.02$
& 2.50 $\pm0.02$
& 2.55 $\pm0.01$
& 2.53 $\pm0.02$
& 2.50 $\pm0.02$
& 2.53 $\pm0.01$
& 2.60 $\pm0.01$ 
\\\\
& $f_{sc}$ & 0.232 $_{- 0.005 }^{+ 0.010 }$
& 0.20 $\pm0.01$
& 0.20 $\pm0.01$
& 0.24 $\pm0.01$
& 0.22 $_{- 0.01 }^{+ 0.02 }$
& 0.22 $\pm0.02$
& 0.23 $\pm0.01$
& 0.23 $\pm0.01$
\\\\
& L$^{(2)}_{Simpl}$/L$_{Edd}$ & 0.302 $\pm0.004$
& 0.184 $\pm0.003$
& 0.201 $\pm0.003$
& 0.241 $\pm0.005$
& 0.220 $\pm0.004$
& 0.126 $\pm0.004$
& 0.207 $\pm0.004$
& 0.277 $\pm0.005$
\\\\
\textit{Nthcomp} & $\Gamma_{Nthcomp}$ & 1.89 $_{- 0.04 }^{+ 0.06 }$
& 1.88 $_{- 0.1 }^{+ 0.08 }$
& 1.96 $\pm0.08$
& 1.71 $_{- 0.13 }^{+ 0.11 }$
& 1.91 $_{- 0.15 }^{+ 0.11 }$
& 1.59 $_{- 0.19 }^{+ 0.17 }$
& 1.55 $_{- 0.15 }^{+ 0.13 }$
& 1.88 $_{- 0.11 }^{+ 0.08 }$
\\\\
& kT$_e$ (eV) & 158.07 $_{- 4.25 }^{+ 4.04 }$
& 160.77 $_{- 5.40 }^{+ 5.41 }$
& 164.07 $_{- 5.32 }^{+ 5.80 }$
& 151.15 $_{- 5.21 }^{+ 5.27 }$
& 155.81 $_{- 7.49 }^{+ 6.41 }$
& 150.31 $_{- 7.19 }^{+ 8.01 }$
& 143.50 $_{- 4.70 }^{+ 5.02 }$
& 155.69 $_{- 5.55 }^{+ 4.80 }$
\\\\
& kT$_{bb}$ (eV)& $<30.03$
& $<29.81$
& $<188.56$
& 33.12 $_{- 2.52 }^{+ 1.91 }$
& 32.46 $_{- 2.26 }^{+ 2.03 }$
& 33.08 $_{- 2.55 }^{+ 1.86 }$
& 34.97 $_{- 1.93 }^{+ 1.58 }$
& $<31.91$
\\\\
& L$^{(3)}_{Nthcomp}$/L$_{Edd}$ & 0.284 $_{- 0.006 }^{+ 0.005 }$
& 0.195 $_{- 0.005 }^{+ 0.004 }$
& 0.186 $\pm0.005$
& 0.236 $\pm0.004$
& 0.200 $\pm0.004$
& 0.135 $\pm0.003$
& 0.216 $\pm0.004$
& 0.261 $\pm0.005$
\\\\
\textit{zgauss} & E$_{Line}$ (keV) & 6.80 $_{- 0.24 }^{+ 0.25 }$
& 6.61 $\pm0.15$
& 6.69 $_{- 0.21 }^{+ 0.20 }$
& 6.77 $_{- 0.14 }^{+ 0.15 }$
& 6.66 $_{- 0.18 }^{+ 0.19 }$
& 6.67 $\pm0.16$
& 6.55 $\pm0.13$
& 6.41 $_{- 0.23 }^{+ 0.22 }$
\\\\
& F$^{(4)}_{Line}$ (10$^{-5}$)& 2.58 $_{- 0.68 }^{+ 0.66 }$
& 2.90 $\pm0.62$
& 2.69 $\pm0.71$
& 3.33 $_{- 0.62 }^{+ 0.63 }$
& 2.88 $_{- 0.66 }^{+ 0.68 }$
& 2.7 $\pm0.6$
& 3.36 $\pm0.62$
& 2.86 $_{- 0.69 }^{+ 0.71 }$
\\
\hline
\\\\
\multicolumn{2}{c}{$\chi^2$/dof~~~~~~~~~~~~~~} & \multicolumn{1}{c}{307.78/251}
& \multicolumn{1}{c}{375.10/248}
& \multicolumn{1}{c}{358.36/247}
& \multicolumn{1}{c}{357.22/250}
& \multicolumn{1}{c}{324.17/248}
& \multicolumn{1}{c}{277.84/238}
& \multicolumn{1}{c}{359.74/249}
& \multicolumn{1}{c}{349.05/248}
\\\\
\multicolumn{2}{c}{$\chi^2_{\nu}$~~~~~~~~~~~~~~} & \multicolumn{1}{c}{1.24} & \multicolumn{1}{c}{1.51} & \multicolumn{1}{c}{1.45} & \multicolumn{1}{c}{1.43} & \multicolumn{1}{c}{1.31} & \multicolumn{1}{c}{1.17} & \multicolumn{1}{c}{1.44} & \multicolumn{1}{c}{1.41} \\
\hline
\end{tabular}
\label{tab_spec}
\end{center}
\begin{flushleft}
~~~~~~~~~~~~~~$^{(1)}${Maximum optical depth for absorption at the threshold energy.}\\
~~~~~~~~~~~~~~$^{(2)}${Normalized hard X-ray luminosity.}\\
~~~~~~~~~~~~~~$^{(3)}${Normalized soft X-ray luminosity.}\\
~~~~~~~~~~~~~~$^{(4)}${Line flux or the normalization of the redshifted gaussian line in units of photons cm$^{-2}$ s$^{-1}$}\\

\end{flushleft}
\end{table}
\end{landscape}

\begin{figure}
\begin{center}
\includegraphics[trim=0cm 0cm 0cm 0.0cm, clip=true, width =8.7cm, angle=-90]{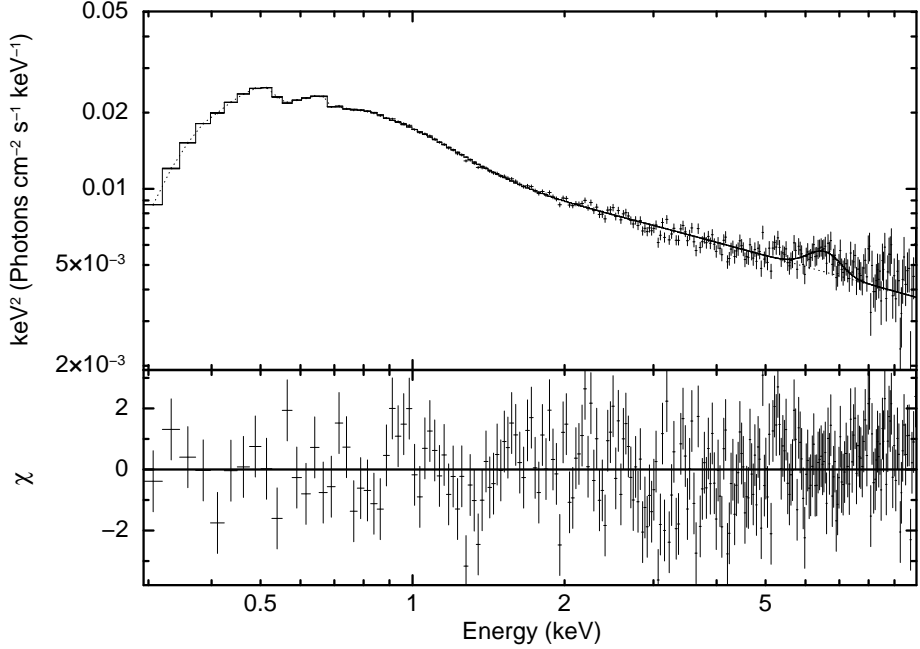}
\caption{Top panel: The Unfolded EPIC-pn spectral data and the best-fitting model \textit{zedge$\times$zedge$\times$TBabs}(\textit{Simpl$\otimes$Nthcomp+zgauss}) for Observation Number 2. Bottom panel: The deviation of the observed data from the model.}
\label{Fig_pn}
\end{center}
\end{figure}

Extrapolation of the model down to 0.3~keV provided a poor fit
indicating the presence of soft excess in the spectrum. We attempted
to describe this soft excess with the thermal Comptonization XSPEC model
\textit{Nthcomp}. In this model, the temperature $kT_{bb}$ of 
seed photons from the accretion disc (blackbody or disc blackbody) parameterizes 
the low energy cut-off, while the high energy roll over is given by the electron 
temperature $kT_{e}$ ($\sim$ 160~eV). We assume that the photons from this component
are the seed photons to the second thermal component giving rise to the high
energy power law emission. Hence, we re-analysed the data replacing the
\textit{powerlaw} by introducing the convolution XSPEC model (\textit{Simpl})
which transfers a fraction of the seed photons in the input spectrum into a
power law \citep{Steiner09}. The model yielded an unacceptable 
fit statistic of $\chi^2$/dof=547.48/253. Inspection of the residuals revealed 
that the soft part of the spectrum was affected by an absorption feature around 
0.7~keV. A better $\chi^{2} = 385.74$ for 251 dof was obtained for the same observation, 
when we fitted the region with a red-shifted absorption edge model, \textit{zedge}. 
Another absorption feature was also found at $\sim$0.5~keV and it was modelled using 
one more \textit{zedge}. Correspondingly the $\chi^{2}$ was improved by $\Delta\chi^{2}=-9.92$ 
which clearly shows the significance of including the new absorption feature. 
We also tried to model these features using more complex models such as \textit{zxipcf} 
and \textit{grid22soft} and obtained fit statistics comparable to the 
phenomenological two edge model. Moreover, the relevant spectral parameters 
such as the hard and soft X-ray fluxes and high energy spectral index are not 
sensitive to the absorption model used. Thus we proceed with the simple 
phenomenological model of two edges in this work. For this model, 
we show the unfolded spectrum and residuals in Fig.~\ref{Fig_pn}.

In order to study the relationship between X-ray and UV emission, we
need to fit the spectra simultaneously. For this we loaded the
\textit{UVW2} data along with the pn data and then added the model
\textit{ezdiskbb} which describes the accretion disc spectrum
consisting of multiple blackbody components \citep{Zimmerman05}. The
model is defined by two parameters, the inner disc temperature
$kT_{in}$ and the norm N$_{disk}$, where N$_{disk}$ is determined by
the inner disc radius and the inclination of the disc. Further the
effect of interstellar extinction on the source spectrum was taken
into account using the model \textit{uvred}, based on Seaton's law
\citep{Seaton79}. This UV reddening model is valid only in the range
1000$-$3074~\AA, and can be used in combination with photoelectric 
absorption models. The parameter E(B-V) was determined from the
reddening law R$_{V}=$ A$_{V}$/E(B-V) \citep{Fitzpatrick99} and the
A$_{V}$ magnitude of 0.198 was obtained from original SFD98 values
assuming R$_{V}=3.1$. So the value of E(B-V) was fixed to 0.064 for
all observations. Then we fitted the \textit{UVW2} data
simultaneously with the EPIC-pn data by tying the parameter $kT_{in}$
to the $kT_{bb}$ of \textit{Nthcomp}. This left only the normalisation
of the \textit{ezdiskbb} model as a free parameter to fit the single
UV data point. To ensure that only the outer disc emission is used to
fit the UV data point, we fixed the normalisation of the Comptonization
component at a negligible value for the UV part of the spectrum.
The best fit parameters for this model for all the observations are
given in Table~\ref{tab_spec}.

We have also calculated the X-ray flux for each model component in the
range 0.3$-$10~keV using the XSPEC convolution model \textit{cflux}. The
flux corresponding to the model component \textit{Simpl} was calculated
from the unabsorbed X-ray flux and the \textit{Nthcomp} flux. It was
obtained by subtracting $(1-f_{sc})$ times the \textit{Nthcomp} flux from
the unabsorbed flux. Using the luminosity distance of \textit{D}=98.5Mpc, by
assuming the cosmological parameters H$_0$=73 km s$^{-1}$ Mpc$^{-1}$,
$\Omega_{m}$=0.27 and $\Omega_\Lambda$=0.73, we obtained the luminosities
of each component.

The normalisation of the \textit{ezdiskbb} model, \textit{N$_{disc}$}
is related to the inner disc radius $R_{in}$ by
\begin{equation}
R_{in}(km)=f^2 \left(\dfrac{N_{disc}}{\cos i}\right)^{1/2} D_{10kpc},
\end{equation} 
where $D_{10kpc}$ is the luminosity distance to the source in units
of 10 kpc. \textit{i} is the inclination angle assumed here to be 30$^{\circ}$
and \textit{f} stands for the colour correction factor which we assume
to be the generally accepted value of 1.7. The mass accretion rate
can be obtained using \citep{Zimmerman05}
\begin{equation}
 \dot{M}=\frac{8\pi\sigma}{3GM} \left(\frac{T_{\ast}}{f}\right)^4
 R_{in}^3 ,
\end{equation}
where $T_{\ast} = T_{in}/0.488$. Several previous studies have estimated the
black hole mass of Ark~564 to be in the range ($1.15-10$)$\times10^6M_\odot$
(e.g. \cite{Pounds01}, \cite{Wang_Lu01}, \cite{Bian_zhao03}, \cite{Botte04},
\cite{Zhou_Wang05}, \cite{Zhang_Wang06ApJ_653_137}). In this study, we adopt
the value 2.61$\times10^6$ M$_\odot$ \citep{Botte04} obtained from stellar velocity
dispersions. This allows us to express the luminosities in terms of the Eddington
Luminosity $L_{Edd}$ and the accretion rate in terms of the Eddington accretion rate, $\dot{M}_{Edd}=\frac{L_{Edd}}{\eta c^2}$, where $\eta$ is the efficiency factor taken
as 0.1. The normalised luminosities for the different observations are listed 
in Table~\ref{tab_spec}.

%======================================================================================================

\begin{table}
\begin{center}
\caption{Reduced $\chi^2$ values for constant model fits to spectral parameters derived from eight observations}
\begin{tabular}{llr}
\hline
\hline
\\
Model & Parameter & $\chi^2_{\nu}$ \\
\hline
\\ 
\textit{zedge} (at $\sim$0.7 keV) & E$_{edge}$ & 0.76 \\
& $\tau_{max}$ & 1.16			\\
\textit{zedge} (at $\sim$0.5 keV)  &  E$_{edge}$ & 2.69	\\
& $\tau_{max}$ & 1.18			\\
\textit{ezdiskbb} & N$_{disc}$ & 0.12 	\\
\textit{Simpl} & $\Gamma_{Simpl}$ & 6.13 	\\
& $f_{sc}$ & 1.53 		\\
\textit{Nthcomp} & $\Gamma_{Nthcomp}$ & 1.58 	\\
& kT$_e$ & 1.56 			\\
& kT$_{bb}$ & 0.23 		\\
& N$_{Nthcomp}$ & 26.07 			\\
\textit{zgauss} &  E$_{Line}$ & 0.42 	\\
&  F$_{Line}$ & 0.21 			\\
Flux & F$_{UV}$ & 61.79 		\\
Luminosity & $L_{Nthcomp}$ & 129.15 \\
& $L_{Simpl}$ & 200.41 \\
\hline
\end{tabular}
\label{tab_var}
\end{center}
\end{table}

\begin{table}
\begin{center}
\caption{Spearman's correlation between different parameters. Column 3: Spearman's rank-order correlation coefficient. Column 4: p-value}
\begin{tabular}{llrr}
\hline
\hline
\\
Parameter 1          & Parameter 2    & r~~ & p~~  \\
\hline
\\
F$_{UV}$ & Count Rate $_{X-ray}$ & 0.69 & 0.06 \\ 
F$_{UV}$ & L$_{Nthcomp}$/L$_{Edd}$ & 0.71 & 0.05 \\
F$_{UV}$ & L$_{Simpl}$/L$_{Edd}$ & 0.57 & 0.14 \\
F$_{UV}$ & $\Gamma_{Simpl}$ & 0.93 & 0.0009 \\
L$_{Nthcomp}$/L$_{Edd}$ &  L$_{Simpl}$/L$_{Edd}$ & 0.95 & 0.0003 \\
L$_{Nthcomp}$/L$_{Edd}$ &  $\Gamma_{Simpl}$ & 0.78 & 0.02 \\
L$_{Simpl}$/L$_{Edd}$ & $\Gamma_{Simpl}$ & 0.69 & 0.06 \\
\hline
\end{tabular}
\label{tab_corr}
\end{center}
\end{table}

\begin{figure*}
\includegraphics[trim=0cm 0cm 0cm 0.0cm, clip=true, width =15.0cm, angle=0]
{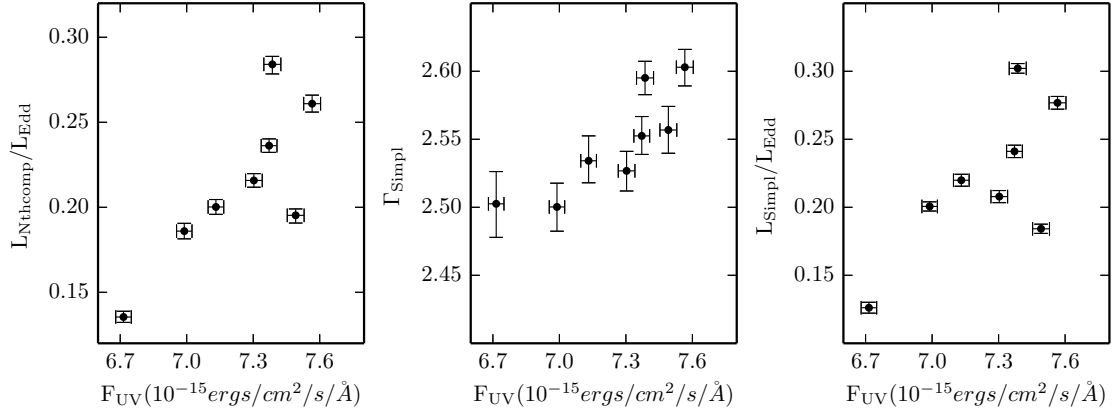}
\caption{The variation of different parameters with UV flux $F_{UV}$. The left and right panels respectively show the soft and the hard X-ray luminosities as a function of $F_{UV}$. The middle panel depicts the dependence of $\Gamma_{Simpl}$ on $F_{UV}$. The luminosities are expressed relative to the Eddington value.}
\label{fig_mdot}
\end{figure*}

\begin{figure*}
\includegraphics[trim=0cm 0cm 0cm 0.0cm, clip=true, width =15.0cm, angle=0]
{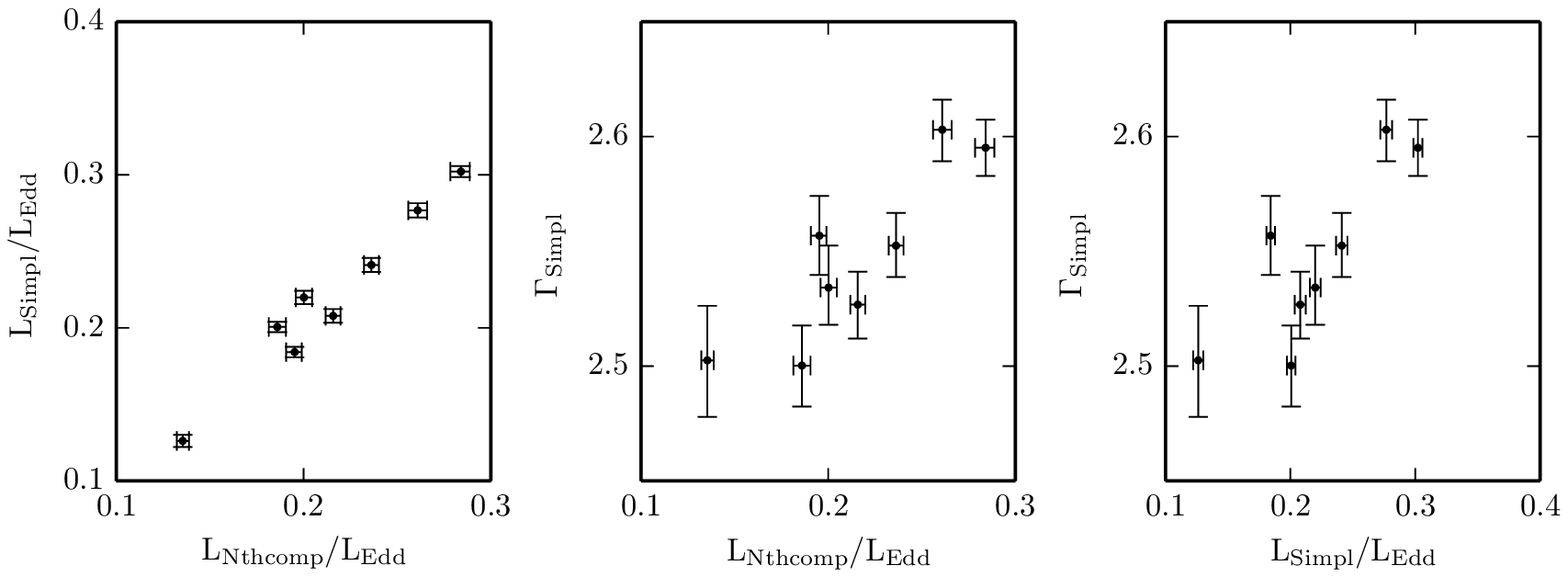}
\caption{The correlation of different parameters. The luminosities are denoted in units of Eddington luminosity. Left $\&$ Middle panels: The variation of hard X-ray luminosity and hard X-ray photon index with soft X-ray luminosity. Right Panel: The dependence  of hard X-ray photon index on the hard X-ray luminosity.}
\label{fig_corr}
\end{figure*}

\section{UV \& X-ray Correlations}

We investigate the correlation between varying X-ray spectral parameters,
luminosities and $F_{UV}$. We identify the variable parameters by fitting
a constant to the best-fit values obtained from different observations
using $\chi^2$ analysis. The reduced  $\chi^2$ values are listed in
Table~\ref{tab_var}, with the larger reduced $\chi^2$ values corresponding
to highly variable parameters. Apart from the luminosities 
of the X-ray components and the UV flux, the high energy photon index $\Gamma_{Simpl}$ 
is also found to be variable. Hence we restrict our correlation analysis to these parameters.

Since the nature of the relations between the parameters are unknown
we need a non-parametric method to calculate the correlation. So we
use Spearman's rank-order correlation \citep{numerical_recipes} to
reveal any correlations. The rank and significance of correlations for
different parameters are listed in Table~\ref{tab_corr}.
Fig.~\ref{fig_mdot} and Fig.~\ref{fig_corr} show the plots of
different parameters for which we find significant correlations.

In the X-ray domain, the soft excess luminosity $L_{Nthcomp}$ is well correlated to 
the high energy power law luminosity $L_{Simpl}$. In fact it seems that 
$\Gamma_{simpl}$ is better correlated with $L_{Nthcomp}$ than it is with the high energy 
luminosity $L_{Simpl}$, in the sense that the null hypothesis probability $p$ is 
significantly smaller. As we discuss in the next section these correlations are consistent 
with the double Comptonization model used for the spectral fitting.

For the UV$-$X-ray correlations, we note that there is not much evidence 
for any correlation between $F_{UV}$ and the X-ray luminosities in the 
two Comptonization component $L_{Nthcomp}$ and $L_{Simpl}$ with probabilities 
of $p = 0.05$ and $0.14$. However a strong correlation is seen for 
$F_{UV}$ and $\Gamma_{simpl}$ with $p=0.0009$.

\section{Summary and Discussion}
We have analysed the simultaneous X-ray (EPIC-pn) and UV (OM
\textit{UVW2}) data from eight \textit{XMM-Newton} observations of
Ark~564 taken in 2011. We used the thermal Comptonization models
favoured by \cite{GCD_Ark564_07} for the spectral fitting of the X-ray
data. The soft X-ray spectrum (0.3$-$3~keV) was modelled by
\textit{Nthcomp} assuming that the UV photons emitted from the
accretion disc are Comptonized by the optically thick corona leading
to the soft excess emission. The accretion disc emitting the seed
photons is described by the multicolour black body model
\textit{ezdiskbb}. Further, the hard X-ray power law emission
(3$-$10~keV) is taken into account by the second Comptonization model
\textit{Simpl} which incorporates the physics of Compton upscattering
of soft photons by hot coronal electrons.  We report that this
double Comptonization model fits all the eight spectra well.

In the X-ray band, we find that the luminosity of the soft Comptonization
component $L_{nthcomp}$ correlates well with the hard component $L_{Simpl}$.
While there is some evidence that the high energy index 
$\Gamma_{Simpl}$ correlates well with both luminosities, there is a seemingly 
stronger correlation between $\Gamma_{Simpl}$ and $L_{nthcomp}$ in the sense 
that null hypothesis probability is smaller. There have been several studies which have shown that the high energy index is correlated with the X-ray flux (e.g.\cite{2002Dewangan}, \cite{vaughan_edelson2001}, \cite{perola1986}). Recently, \citealt{sarma2015} have studied the index versus flux variation for
Mrk~335 and Ark~564 and have reported that while the correlation exists for
both sources, there is significantly more scatter for Ark~564. Our results
are broadly consistent with their finding and perhaps gives an explanation
for the difference between the two sources. Also the correlations obtained 
here are consistent with the double Comptonization model used.

There is little evidence for any correlation between the UV flux and the 
hard component luminosity with null hypothesis probability of $p = 0.14$. 
There is a hint of a correlation between the UV flux and the soft X-ray 
component luminosity with $p = 0.05$. However, the UV flux is 
strongly correlated to the photon index $\Gamma_{Simpl}$.

We can interpret the correlation between UV and X-ray emissions in two
different ways. One interpretation is that the variation in UV
emission could be due to the accretion rate fluctuation. So as the UV
flux varies, it provides a way to measure the accretion rate $\dot{m}$ as a
fraction of the Eddington rate and it is found to vary from
$\sim$3.7 to $\sim$4.4. However, one can estimate from the accretion rate that 
the measured UV emission should mostly arise from the outer disc at a distance of 
$\sim$~770$R_g$, where $R_g$ is the Schwarzschild radius. At this radius the
viscous time-scale is $t_{visc}\sim9$~years, which is much longer than
the 6~day variability seen; more importantly, the variation
in accretion rate could not have propagated to the inner regions
in such short time-scales. We can further estimate that less than
2\% of the UV flux would arise from radii $\sim$30$R_g$. If
the flux variation at those radii is very large, it may give rise
to the $\sim$1\% variation seen in the UV. However, even at
$\sim$~30$R_g$ the viscous time-scale is too long at $\sim$~61~days.
Moreover, the accretion rate inferred from this interpretation is 
significantly higher than the Eddington rate and hence unlikely. 
Thus, it seems that the UV emission, or at least its variability
cannot arise due to accretion rate fluctuations.

The second possibility is that the UV flux variation is due to the
reprocessing of X-rays. In such a case our results indicate 
that the soft X-ray emitting region and $\Gamma_{Simpl}$ are more important 
in determining the X-ray irradiation than the X-ray luminosity itself. 
We note that recently \cite{2016MNRAS.457..875P} have studied the UV-X-ray correlation 
on much shorter 20~ksec time-scales and have come to a similar conclusion that 
the geometry of the inner X-ray producing region may be playing an important role 
in determining the UV emission.

Though we have not used the blurred reflection model (e.g., \citealt{Fabianetal2002})
to describe the soft X-ray excess and the broad iron line observed from Ark~564,
the model can be tested against the observed correlations. In the blurred reflection
model, the soft excess and the broad iron line are physically the same spectral component,
and hence these two features must be strongly correlated. The presence of the broad iron line
strongly suggests some contribution of the blurred reflection to the soft X-ray excess.
Indeed, the observation of reverberation soft lags of $\sim 100~s$ in Ark~564 by \cite{Kara13}
clearly demonstrate the presence of blurred reflection in the soft ($0.3-1{\rm~keV}$) band.
However, the findings of soft leads in Ark~564 \citep{2002Dewangan,Kara13} suggest contribution
of additional spectral component in the soft band. In our analysis, the broad iron line
does not appear to follow the strongly variable soft X-ray excess emission. Illumination
of the hard X-ray power law component should not only result in the blurred reflection (the
soft excess, broad iron line and the hump in the $\sim20-40{\rm~keV}$) but also in the
reprocessed emission in the UV band. The correlation between $F_{UV}$ and $L_{Simpl}$
may result from the reprocessing of the coronal X-ray emission in the disc. However,
the similar variability amplitudes of the soft excess and the hard X-ray emission
is difficult to explain in the reflection model in which a compact corona along
the symmetrical axis illuminates the disc. In such a model, due to the bending of light,
the reflected emission that includes the soft excess and the iron line is much less variable
than the illuminating power law \citep{Fabian_Miniutti2004}. Thus, the observed strong
soft X-ray excess entirely is unlikely to be the reflected emission.

Our results are based on eight observations separated by $\sim$6~days
and clearly there is a need for a larger number of such observations
to verify these interpretations. Further long term simultaneous monitoring 
of UV and X-ray emissions over different time-scales can give us more insight 
into the variability of the source. Also the results can be compared with
correlations obtained for other AGN. This may be possible with
\textit{ASTROSAT} which has the Ultra Violet Imaging Telescope
(\textit{UVIT}) for monitoring the UV emission and the Soft X-ray
imaging Telescope (\textit{SXT}) and the Large Area X-ray Proportional
Counters (\textit{LAXPC}) for X-ray studies. \\\\

%//

\section*{Acknowledgments}

This study is based on the observations obtained with \textit{XMM-Newton} satellite,
an ESA science mission with instruments and contributions directly funded by ESA
Member States and the USA (NASA). This work has made use of the NASA/IPAC
Extragalactic Data base which is operated by the Jet Propulsion Laboratory,
California Institute of technology and data obtained through the High Energy
Astrophysics Science Archive Research Center Online Service, provided by the NASA/GSFC. 
The first author would like to thank the Department of Science and Technology 
for the support grant (No.SR/S2/HEP-07/2012) for this work.

\def\aj{AJ}%
\def\actaa{Acta Astron.}%
\def\araa{ARA\&A}%
\def\apj{ApJ}%
\def\apjl{ApJ}%
\def\apjs{ApJS}%
\def\ao{Appl.~Opt.}%
\def\apss{Ap\&SS}%
\def\aap{A\&A}%
\def\aapr{A\&A~Rev.}%
\def\aaps{A\&AS}%
\def\azh{AZh}%
\def\baas{BAAS}%
\def\bac{Bull. astr. Inst. Czechosl.}%
\def\caa{Chinese Astron. Astrophys.}%
\def\cjaa{Chinese J. Astron. Astrophys.}%
\def\icarus{Icarus}%
\def\jcap{J. Cosmology Astropart. Phys.}%
\def\jrasc{JRASC}%
\def\mnras{MNRAS}%
\def\memras{MmRAS}%
\def\na{New A}%
\def\nar{New A Rev.}%
\def\pasa{PASA}%
\def\pra{Phys.~Rev.~A}%
\def\prb{Phys.~Rev.~B}%
\def\prc{Phys.~Rev.~C}%
\def\prd{Phys.~Rev.~D}%
\def\pre{Phys.~Rev.~E}%
\def\prl{Phys.~Rev.~Lett.}%
\def\pasp{PASP}%
\def\pasj{PASJ}%
\def\qjras{QJRAS}%2215.bib
\def\rmxaa{Rev. Mexicana Astron. Astrofis.}%
\def\skytel{S\&T}%
\def\solphys{Sol.~Phys.}%
\def\sovast{Soviet~Ast.}%
\def\ssr{Space~Sci.~Rev.}%
\def\zap{ZAp}%
\def\nat{Nature}%
\def\iaucirc{IAU~Circ.}%
\def\aplett{Astrophys.~Lett.}%
\def\apspr{Astrophys.~Space~Phys.~Res.}%
\def\bain{Bull.~Astron.~Inst.~Netherlands}%
\def\fcp{Fund.~Cosmic~Phys.}%
\def\gca{Geochim.~Cosmochim.~Acta}%
\def\grl{Geophys.~Res.~Lett.}%
\def\jcp{J.~Chem.~Phys.}%
\def\jgr{J.~Geophys.~Res.}%
\def\jqsrt{J.~Quant.~Spec.~Radiat.~Transf.}%
\def\memsai{Mem.~Soc.~Astron.~Italiana}%
\def\nphysa{Nucl.~Phys.~A}%
\def\physrep{Phys.~Rep.}%
\def\physscr{Phys.~Scr}%
\def\planss{Planet.~Space~Sci.}%
\def\procspie{Proc.~SPIED}%
\let\astap=\aap
\let\apjlett=\apjl
\let\apjsupp=\apjs
\let\applopt=\ao

\bibliographystyle{raa} % (uses file "plain.bst")
\bibliography{bibtex} % expects file "myrefs.bib"

\end{document}